\documentclass[12pt,preprint]{aastex}

\newcommand{\etal}{\emph{et al.}}

\shorttitle{Superluminal problem in diffusion} %
\shortauthors{V. Berezinsky \& A. Gazizov}

\begin{document}

\title{The problem of superluminal diffusion of relativistic particles
and its phenomenological solution}

\author{R. Aloisio, V. Berezinsky and A. Gazizov}

\affil{INFN, Laboratori Nazionali del Gran Sasso, I--67010
Assergi (AQ), Italy}

\begin{abstract}
We discuss the problem of superluminal propagation in diffusion of
ultra-high energy protons with energy losses taken into account. The
phenomenological solution of this problem is found with help of
generalized J\"{u}ttner propagator, originally proposed for the
relativization of Maxwellian gas distribution. It is demonstrated
that the generalized J\"{u}ttner propagator gives correct
expressions in limits of diffusive and rectilinear propagation of
the charged particles in magnetic fields, together with the
intermediate regime, in all cases without superluminal velocities.
This solution, very general for the diffusion, is considered for two
particular cases: the diffusion inside stationary objects, like
e.g.\ galaxies, clusters of galaxies etc, and for the expanding
universe. The comparison with previously obtained solutions for
propagation of UHE protons in magnetic fields is performed.
\end{abstract}


\section{Introduction}
\label{Introduction}%
Diffusion equation describes many physical processes: thermal
conductivity, the Brownian motion, the viscous motion in gases,
propagation of charged particles in magnetic field and others. Our
main interest is the diffusion of charged relativistic particles
(cosmic rays). For collisional processes in gas, the diffusion
equation can be obtained from kinetic equations \citep{Pitaev}. The
derivation of diffusion equation for motion of charged particles in
magnetic fields is given in \citep{BBDGP}.

In the extreme case, neglecting energy losses of particles and
convection, assuming the diffusion coefficient $D(t,\vec{r})$
independent of the time and space coordinates, the diffusion
equation for the space density of particles $n(\vec{r},t)$ has a
simple form:
\begin{equation}
\label{eq:deq1}%
  \frac{\partial}{\partial t}n(\vec{r},t) - D \nabla^2
n(\vec{r},t) = Q(\vec{r},t),
\end{equation}
where $Q(\vec{r},t)$ is the generation function. The diffusion
coefficient is defined phenomenologically through the flux density
of particles as
\begin{equation}
\label{flx}%
   \vec{j}(\vec{r},t)= - D \nabla n(\vec{r},t).
\end{equation}
Both diffusion equation (\ref{eq:deq1}) and definition of the flux
density do not know about light speed $c$ and in fact result in
superluminal motion. A common example demonstrating it (see e.g.\
\citep{Dunkel-prd}) can be described as follows. From
Eq.~(\ref{flx}) one can find the streaming velocity
$\vec{u}(\vec{r},t)$, determined from $\vec{j}=n\vec{u}$, as
\begin{equation}
\label{vecu}%
   \vec{u} = -D \nabla n/n.
\end{equation}

For the stationary ($\partial n / \partial t = 0$) spherically
symmetric diffusion one has from Eq.~(\ref{eq:deq1})
\begin{equation}
\label{ndndr}%
  n = \frac Q{4\pi r D} \;\;\; \mbox{and} \;\;\; \frac{\partial n}{\partial
   r} = -\frac Q{4\pi r^2 D},
\end{equation}
and finally, using the definition of the diffusion length $l_d$ from
$D \sim c l_d$ for diffusion of relativistic particles with
velocities $v \sim c$, one obtains from (\ref{vecu}) and
(\ref{ndndr})
\begin{equation}
\label{u}%
   u \sim c\; \frac{l_d}r,
\end{equation}
which at $r < l_d$ results in $u>c$.

Another way to demonstrate the problem of superluminal propagation
in diffusion is given by consideration of the average displacement
of a particle in the diffusive regime, $r^2 \sim D t$. One obtains
for average velocity of displacement $v \sim r/t \sim c (l_d/r)$
with the same problem as above. From now on we shall refer to this
problem as superluminal diffusion.

From the practical point of view the described problem of
superluminal diffusion, widely discussed in the literature, can be
in many cases easily avoided. In particular, for diffusion of
charged particles in magnetic fields, a particle deflects at length
$l_d$ on average by angle $\theta \sim 1$ and the movement of a
particle at $r \lesssim l_d$ can be considered as quasi-rectilinear
as long as the energy spectra are concerned and as we did in our
works \citep{BG06,BG07}. In these works, discussing the diffusion of
ultra-high energy cosmic ray (UHECR) protons, we have met a more
serious problem of superluminal propagation connected with inclusion
of proton energy losses.

The diffusion equation for protons with a source at $\vec r
= \vec{r}_g$, energy losses in the form $dE/dt = -b(E)$ and with
time-independent diffusion coefficient $D(E)$, reads
\begin{equation}
\label{eq:DiffEqn}%
\frac{\partial}{\partial t}
n_p(E,\vec{r},t)-D(E)\nabla^2n_p(E,\vec{r},t) -
\frac{\partial}{\partial E} \left[ b(E) n_p(E,\vec{r},t)  \right] =
Q(E,t)\delta^3(\vec{r} - \vec{r}_g).
\end{equation}
The solution of this equation was found by Syrovatsky
\citep{Syrovatsky} as
\begin{equation}
\label{SyrovSolution}%
n_p(E,\vec{r},\vec{r}_g)=\frac{1}{b(E)} \int \limits_E^\infty dE_g
Q(E_g)
\frac{\exp\left[-(\vec{r}-\vec{r}_g)^2/4\lambda(E,E_g)\right]}{\left[
4\pi\lambda(E,E_g)\right]^{3/2}},
\end{equation}
where $n_p$ is the space density of relativistic protons, $E$ is the
observed energy, $E_g$ is the generation energy of a proton in the
source, and $\lambda(E,E_g)$ given by
\begin{equation}
\label{lambdaSyrov}%
\lambda(E,E_g)=\int \limits_E^{E_g} dE'\; \frac{D(E')}{b(E')},
\end{equation}
has a meaning of a distance squared traversed by a proton in a given
direction during the time, when the proton energy decreases from
$E_g$ to $E$.

The superluminal diffusion in solution (\ref{SyrovSolution}) is
immediately seen from the fact that it is the solution of
Eq.~(\ref{eq:DiffEqn}) only when the lower limit of integration is
the observed energy $E_g^{{\rm min}} = E$. One may prove it
substituting (\ref{SyrovSolution}) into (\ref{eq:DiffEqn}). In
particular, at any other lower limit $E_g^{\rm min}$ the
delta-function in the r.h.s.\ of Eq.~(\ref{eq:DiffEqn}) is not
reproduced.

Thus, the minimum generation energy at distance $r$ in the solution
of Eq.~(\ref{eq:DiffEqn}) equals to the observed energy $E$ and it
can be interpreted in the unique way: the propagation time $\tau
\rightarrow 0$ and thus the energy loss of a particle is absent. It
implies the velocity of a particle $v \rightarrow \infty$.

\begin{figure}
\begin{center}
\includegraphics[width=0.49\textwidth,angle=0]{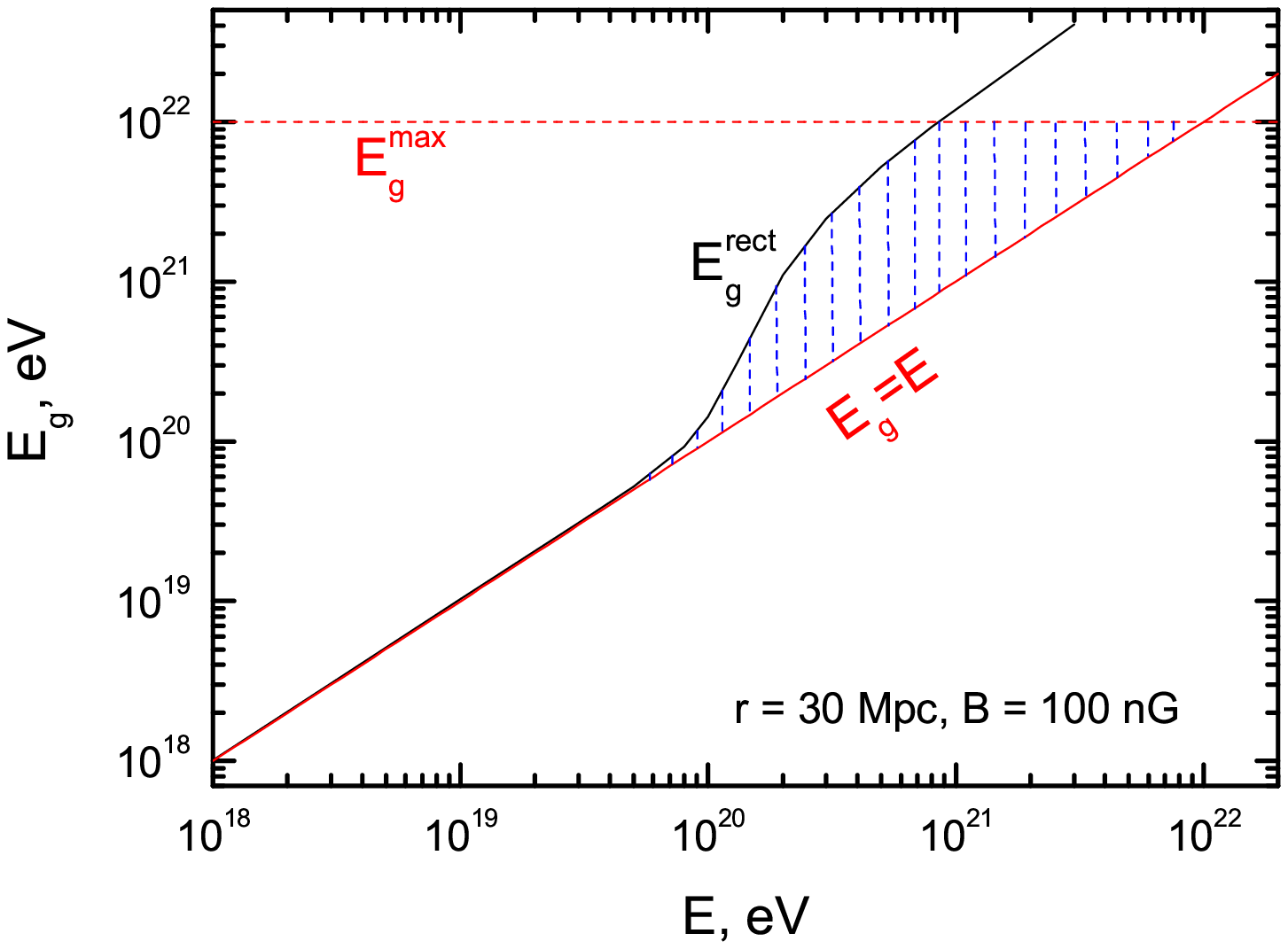}
\includegraphics[width=0.49\textwidth,angle=0]{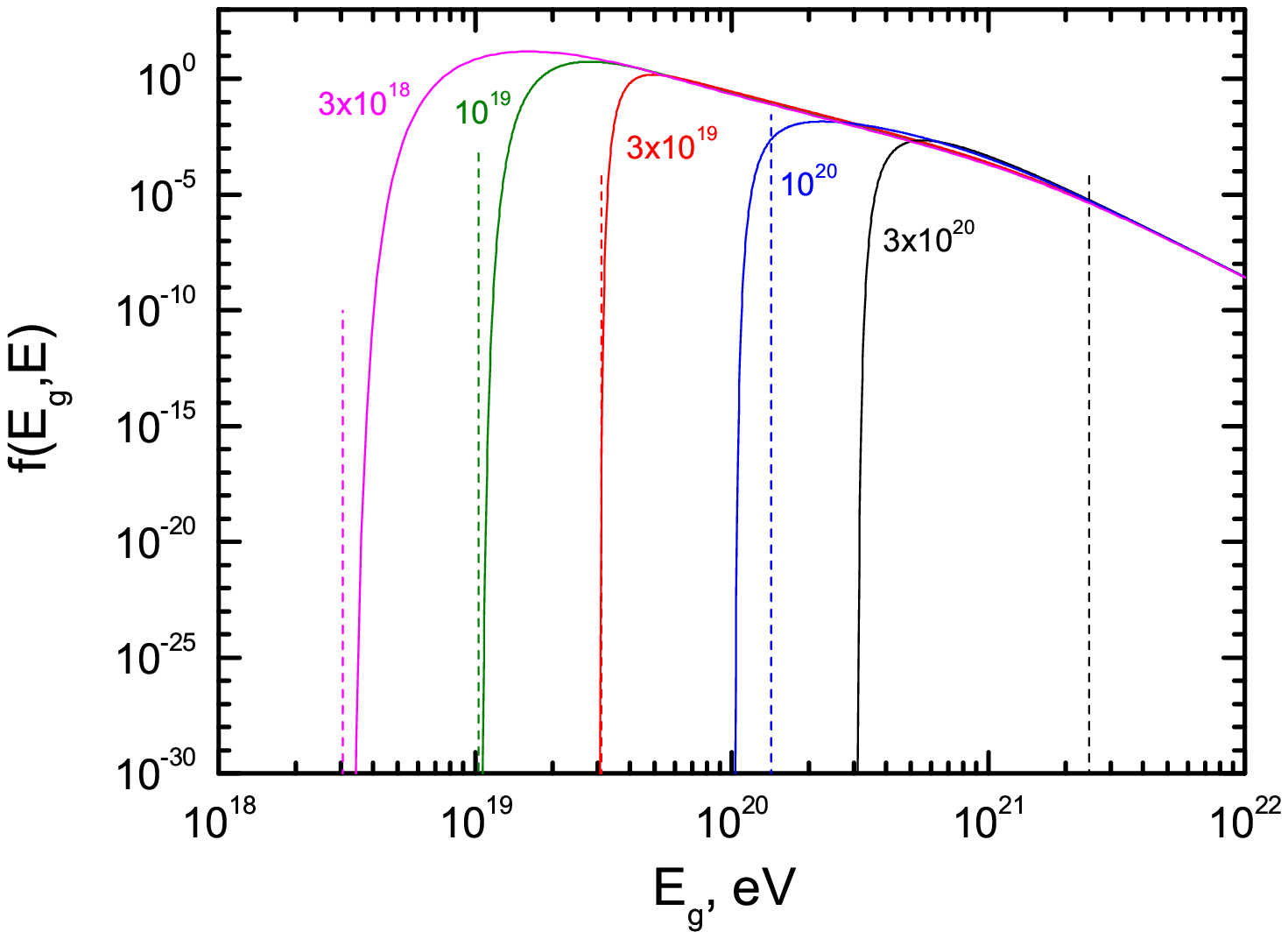}
\end{center}
\caption{
\label{Fig1}%
The superluminal propagation problem in the Syrovatsky solution
caused by energy losses of UHE protons for $B_c=100$~nG and distance
to the source $r=30$~Mpc. In the left panel is shown the region in
($E_g\;,E$) plane (above $E_g=E$ line) allowed by
Eq.~(\ref{SyrovSolution}). The line $E_g = E_g^{{\rm rect}}(E,r)$
corresponds to rectilinear propagation and to the physical lower
limit in the solution (\ref{SyrovSolution}). The hatched region
between $E_g=E$ and $E_g=E_g^{\rm rect}$ corresponds to superluminal
velocities. For $E \lesssim 1\times 10^{20}$~eV the solutions
practically do not have superluminal velocities. In the right panel
the integrand of Eq.~(\ref{SyrovSolution}) is shown as function of
$E_g$ for fixed $E$ with values indicated by the numbers, given in
eV. The dashed lines show the rectilinear physical lower limits
$E_g^{\rm min}$. At $E \lesssim 1\times 10^{20}$~eV the regions
below these limits, i.e. with $v>c$ are negligibly small, while at
$E > 1 \times 10^{20}$ eV they give considerable contribution to Eq.
(\ref{SyrovSolution}).}
\end{figure}

The \emph{physical} value of the minimal generation energy
$E_g^{{\rm min}}(E,r)$ is given by the rectilinear propagation and
it can be easily calculated if $b(E)=-dE/dt$ is known. But with this
$E_g^{{\rm min}}$ Eq.~(\ref{SyrovSolution}) is not any more the
solution of Eq.~(\ref{eq:DiffEqn}). In the left panel of
Fig.~\ref{Fig1} the unphysical region of the solution
(\ref{SyrovSolution}) with superluminal velocity is shown as the
hatched area.

\section{The problem of superluminal diffusive propagation of UHECR
and how it has been eliminated}
\label{SuperluminalProblem}%
From the formal mathematical point of view the superluminal
propagation exists always in the solutions of the non-relativistic
diffusion equations, but very often the contribution of unphysical
regions to the solution is negligibly small (e.g.\ in heat
conductivity). In these cases one may say that diffusion equation
gives the correct description of the considered physical
phenomenon. It is similar to validity of the Maxwell distribution
for molecules in low-temperature gas (this analogy will be
considered in the next section), where formally there are particles
with velocities $v \to \infty$.

In our works \citep{AB04,AB05} on diffusion of UHE protons in static
universe and in the works \citep{BG06,BG07} on diffusion in the
expanding universe we recognized the problem of superluminal
propagation and eliminated it using the following recipes:

(i) For calculation of flux from a source at distance $r\lesssim
l_d(E)$ for protons with energies $E$ we used the rectilinear
propagation. Then the problem of superluminal diffusion disappears,
but a technical problem of sewing the rectilinear and diffusive
solutions arises, since the rectilinear propagation starts at $r
\sim l_d(E)$, while diffusive regime begins at $r\sim 6 l_d(E)$. In
the paper \citep{AB05} we used some smooth interpolation between two
regimes, while in the paper \citep{BG07} we used for transition the
fixed energy $E_{t}(r)$, where rectilinear and diffusive fluxes are
equal. It was done intentionally, because the energy of transition
appeared in the calculated diffuse spectrum as a feature, which
indicates the change of the propagation regime.

(ii) The superluminal propagation is caused also by energy losses of
particles during diffusion, and it appears as $E_g^{\rm min}=E$ in
Eq.~(\ref{SyrovSolution}). This problem is ameliorated by presence
of the {\it magnetic horizon} (see below): the problem of
superluminal diffusion is most severe for the sources at largest
distances beyond the magnetic horizon, while contribution of these
sources to the observed diffuse flux is small.

According to our calculations the picture for propagation of UHE
protons in extragalactic magnetic fields looks as follows. At low
energies diffusion dominates. At high energies, determined by energy
dependence of the diffusion length $l_d(E)$ and by size of horizon
$r_{\rm hor}(E)$, the propagation changes to (quasi)rectilinear,
which has no problem with superluminal propagation. This picture is
valid for magnetic fields with the critical values $B_c$ (see below)
between $0.01 - 10$~nG. For the fields $B_c \gtrsim 100$~nG the
superluminal signal appears at the highest energies in the diffuse
spectrum (see Fig.~1). The fields $B_c \lesssim 1.6\times
10^{-3}$~nG provide quasi-rectilinear propagation for all protons at
energies $E \geq 1\times 10^{17}$~eV, while at lower energies the
diffusion becomes unavoidable \citep{BG07}. For a reasonably low
field $B_c \approx 0.01$~nG, diffusion becomes valid at $E \lesssim
1\times 10^{18}$~eV.

Some technical details and numerical results on the problem of
superluminal diffusion are in order.

We have considered diffusion of UHECR in magnetized turbulent plasma
with random magnetic field, and with the coherent magnetic field
$B_c$ on the basic linear scale $l_c$. It determines the critical
energy $E_c \approx 1\times 10^{18}(B_c/1~{\rm nG})(l_c/1~{\rm
Mpc})$~eV, calculated from relation $r_L(E_c) \approx l_c$, where
$r_L(E)$ is the Larmor radius of particle in magnetic field $B_c$.
The diffusion length $l_d(E) = l_c(E/E_c)^2$ at $E > E_c$ for all
spectra of turbulence and $l_d(E) = l_c(E/E_c)^n$ at $E < E_c$ with
$n=1/3$ and $n=1$ for the Kolmogorov and Bohm diffusion,
respectively. At intermediate energies the interpolation of $l_d(E)$
between these two regimes was used. For low magnetic field $B_c \leq
1$ nG and at $E > E_c$ the diffusion length $l_d(E)$ becomes very
large at high energy and propagation of UHE protons becomes
rectilinear at all highest energies. The superluminal propagation
thus disappears at all energies above the energy of transition
$E_{tr}$ from diffusive to rectilinear regime. In calculation of
$E_{{\rm tr}}$ the \emph{magnetic horizon} is involved
\citep{AB05,Aloisioetal} (for physical discussion of magnetic
horizon see \citep{Parizot}). Magnetic horizon is determined as a
distance traversed by a particle during the age of a universe $t_0$:
\begin{equation}
\label{r2hort}%
r^2_{{\rm hor}} = \int \limits_0^{t_0} dt\; D[E_g(E,t)],
\end{equation}
where $E_g(E,t)$ is the energy that a particle has at time $t$, if
its energy at $t = t_0$ is $E$. Putting $dt = -dE_g/b(E_g)$ in
Eq.~(\ref{r2hort}) we obtain
\begin{equation}
\label{r2horE}%
r^2_{{\rm hor}} = \int \limits_E^{E_{{\rm max}}} dE_g
\frac{D(E_g)}{b(E_g)},
\end{equation}
where $E_{\max} = \min[E_g(E,t_0), E_{\max}^{{\rm acc}}]$, and
$E_{\max}^{{\rm acc}}$ is the maximum acceleration energy.

In $r^2_{{\rm hor}}$ one may recognize the Syrovatsky variable
$\lambda(E,E_g)$ at $E_g = E_{\max}$. The Syrovatsky solution
(\ref{SyrovSolution}) knows about the horizon suppression of
$n(E,r)$ at distance $r\geq \sqrt{\lambda(E,E_g)}$ by exponential
factor $\exp(-r^2/4\lambda)$.

The energy of transition from diffusive to rectilinear propagation
is determined for diffuse fluxes by condition $l_d(E) \gtrsim
r_{{\rm hor}}(E)$, or $l_d^2(E_{{\rm tr}}) \sim \lambda(E_{{\rm
tr}},E_{\max})$ with $E_{\max}$ defined as above.

If magnetic field is very strong, $B_c \sim 100$ nG \citep{AB04},
and hence $E_c \sim 10^{20}$ eV, the diffusive regime continues up
to very high energies and contribution of unphysical solutions (with
$v
> c$) appears. In Fig.~{\ref{Fig1}} the integrand of
Eq.~(\ref{SyrovSolution}) is shown as a function of $E_g$ for $B_c =
100$ nG and distance to a source $r = 30$ Mpc. The unphysical
regions with $v > c$ are negligibly small at $E\leq3\times 10^{19}$
eV, but is large at $E=3\times 10^{20}$ eV. In this case we limited
our consideration by energies $E < 1.2\times 10^{20}$ eV (see Fig.~6
in \citep{AB04}).


\section{Approach to solution of the superluminal propagation problem}
\label{sec:Approach}%
The most radical way to eliminate the superluminal signal in
non-relativistic diffusion equation is given by relativization of
the diffusion equation similar to relativization of the
Schr\"{o}dinger equation in quantum mechanics. However, after more
than 70 years of efforts in this direction no successful solution
has been found (see \citep{Dunkel-prd} for review and references).
In this paper we shall follow another approach first suggested by
F.~J\"{u}ttner \citep{Jutt} for the relativization of the Maxwellian
distribution of particles. We apply this method to the Green
function of the diffusion equation.

In the work \citep{Dunkel-prd} it was observed that the problem of
relativization of the Maxwell distribution is identical to the
relativization of diffusion propagator (the Green function). The
normalized (per unit phase volume) probability density function of
the Maxwell distribution of particles with mass $m$ and temperature
$T$ is given by
\begin{equation}
P_M(v)= \left (\frac{m}{2\pi kT}\right )^{3/2} \exp\left
(-\frac{mv^2} {2kT} \right ). \label{eq:Max}
\end{equation}

Changing $v \to x$ and $kT/m \to 2Dt$ one obtains the Green function
of the diffusion equation (\ref{eq:deq1}):
\begin{equation}
\label{PropagatorSyrovatsky}%
P_{\rm diff}(r,t) = \frac{1}{(4 \pi D t)^{3/2}}\; \exp \left( -
\frac{r^2}{4 D t} \right),
\end{equation}
where $r$ is a distance to the source and $D$ is time-independent
diffusion coefficient.

Therefore, one can use for the diffusive Green function the
J\"{u}ttner distribution, where superluminal velocities are absent.

We start with the phenomenological definition of the propagation
function (propagator). Let us consider first the static universe or
more generically an extended stationary object like a galaxy or a
cluster of galaxies. The propagation of the charged particles in
magnetic fields may be described with help of propagator $P(E,t,r)$,
where $E$ is the observed energy, $t$ is the propagation time and
$r$ is the distance to a source. The definition of the propagator
is given by
\begin{equation}
\label{PropagatorSolution}%
n(E,r) = \int_0^{\infty} dt\; Q[E_g(E,t),t]\; P(E,t,r)\;
\frac{dE_g}{dE}(E,t),
\end{equation}
where $n(E,r)$ is the observed space density of particles,
$E_g(E,t)$ is energy of a particle at time $t$, analytic
expression for $dE_g/dE$ is given in \citep{BGG06}, and $Q(E,t)$ is
the source generation function.

The propagator $P(E,t,r)$ can be thought of as a
Green function of an unknown relativistic equation.

For rectilinear propagation of ultrarelativistic particles with $v
\approx c$ the propagator is given by
\begin{equation}
\label{PropagatorRectilinear}%
P(E,t,r) = \frac{1}{4 \pi c^3 t^2}\; \delta(t-\frac r c)\;,
\end{equation}
and Eq.~(\ref{PropagatorSolution}) results in
\begin{equation}
\label{RectilinearSolution}%
n(E,r) = \frac{Q[E_g(E,r/c),r/c]}{4 \pi c r^2}\; \frac{dE_g}{dE},
\end{equation}
which coincides with the expression obtained from conservation of
number of particles.

For diffusive propagation the propagator can be obtained from the Syrovatsky
solution (\ref{SyrovSolution}) as
\begin{equation}
\label{PropagatorSyrovatskyDiff}%
P(r,t) = \frac{1}{[4 \pi \lambda(E,t)]^{3/2}} \exp \left[ -
\frac{r^2}{4 \lambda(E,t)} \right],
\end{equation}
where
\begin{equation}
\label{lambda}%
\lambda(E,t) = \int_0^t dt' D[E(t')] = \int_E^{E_g(t)} d E'\;
\frac{D(E')}{b(E')}
\end{equation}
and $D(E)$ is the diffusion coefficient; $b(E) = -dE/dt$ is the
energy loss of a particle. The density of particles $n_{\rm
diff}(E,r)$ is given by Eq.~(\ref{PropagatorSolution}).

Both propagation functions $P_{\rm rec}(E,t,r)$ and $P_{\rm
diff}(E,t,r)$ are normalized by unity
\begin{equation}
\label{normalization}%
\int dV P(E,t,r) = 1
\end{equation}
and thus they have a meaning of probability to find a particle in a
unit volume at distance $r$ from a source at time $t$ after
emission. It is easy to see from Eqs.~(\ref{PropagatorRectilinear})
and (\ref{RectilinearSolution}) that $P(E,t,r)$ has the correct
dimension $[P] = L^{-3}$.

The J\"{u}ttner distribution in terms of $r=v$ and $2 D t = k T/m$,
as was described above, has been given in \citep{Dunkelv2} as
\begin{equation}
\label{PropagatorJuettner}%
P_J(E,t,r) = \frac{\theta(c t - r)}{(ct)^3 Z\left(\frac{c^2 t}{
2D}\right) \left[ 1 - \left( \frac{r}{c t} \right)^2 \right]^2} \; %
\exp \left[ -\frac{ \frac{c^2 t}{2D}}{\sqrt{1 - \left(\frac{r}{c
t}\right)^2}} \right],
\end{equation}
where
\begin{equation}
\label{JuettnerNorm}%
Z(y) = 4 \pi K_1(y)/y
\end{equation}
with $K_1(y)$ being the modified Bessel function. One may observe
that the superluminal propagation with $v = r/t > c$ is forbidden
for this propagator.

Instead of the J\"{u}ttner function (\ref{PropagatorJuettner}) we
introduce the generalized J\"{u}ttner function $P_{gJ}(E,t,r)$,
imposing to it two limiting conditions of transition to rectilinear
propagator (\ref{PropagatorRectilinear}) and the Syrovatsky
propagator (\ref{PropagatorSyrovatsky}), and keeping the condition
of subluminal velocities $r \leq ct$. For this purpose we transform
the Eq.~(\ref{PropagatorJuettner}) as follows:
\begin{equation}
\label{substitute}%
\frac{c^2t}{2D} \rightarrow \frac{c^2t^2}{2\lambda[E_g(E,t)]} \equiv
\alpha(t),
\end{equation}
and use instead of $t$ the new variable
\begin{equation}
\label{xi}%
\xi(t) = r/ct,
\end{equation}
where $\lambda[E_g(E,t)]$ is given by Eq.~(\ref{lambda}). Note that
the both new quantities are dimensionless.

The generalization imposed by Eq.~(\ref{substitute}) is motivated by
the time-dependent diffusion coefficient $D[E_g(t)]$ and by the
presence of energy losses. In this case we generalize the implicit
quantity $D \times t$ in the J\"{u}ttner distribution
(\ref{PropagatorJuettner}) to $\int D(t)dt = \lambda(E,t)$ given by
Eq.~(\ref{lambda}). As a result we have
\begin{equation}
\label{change}%
\frac{c^2t}{2D} = \frac{c^2t^2}{2Dt} \rightarrow \frac{c^2t^2}{2\int
D(E,t)dt} = \frac{c^2t^2}{2\lambda(E,t)} \equiv \alpha(E,t).
\end{equation}
In terms of $\xi$ and $\alpha(E,t)$ the generalized J\"{u}ttner
function $P_{gJ}(E,t,x)$ and density of particles $n(E,r)$ are given
by:
\begin{equation}
\label{PropagatorJuettmJ}%
P_{gJ}(E,t,r) = \frac{\theta(1-\xi)}{4 \pi (c t)^3 }\;
\frac{1}{(1-\xi^2)^2}\; \frac{\alpha(E,\xi)}{K_1[\alpha(E,\xi)]}\;
\exp\left[- \frac{\alpha(E,\xi)}{\sqrt{1-\xi^2}} \right],
\end{equation}

\begin{equation}
\label{nJuettmJ}%
n(E,r) = \frac 1{4 \pi c r^2} \int \limits_{\xi_{{\rm min}}}^1 d\xi\;
\frac{Q[E_g(E,\xi)]}{(1-\xi^2)^2} \; \xi\;
\frac{\alpha(E,\xi)}{K_1[\alpha(E,\xi)]}\; \exp\left[-
\frac{\alpha(E,\xi)}{\sqrt{1-\xi^2}} \right] \frac{dE_g}{dE}.
\end{equation}

Now we prove that Eqs.~(\ref{PropagatorJuettmJ}) and
(\ref{nJuettmJ}) have the correct rectilinear and diffusive
asymptotic behavior.

Consider first the high energy regime, where rectilinear propagation
is expected. At $E' > E_c$ the diffusion coefficient increases with
$E'$ as $D(E') \propto (E'/E_c)^2$, while $E'(t)$ increases
exponentially with time \citep{AB04}. As a result $\lambda(E,t)$
increases exponentially with time too, providing small $\alpha$ from
Eq.~(\ref{change}) and (quasi)rectilinear propagation. On the other
hand from Eq.~(\ref{xi}) we have $\xi(t) \approx 1$ as the second
condition of thus (quasi)rectilinear propagation. We impose both of
these conditions to Eq.~(\ref{nJuettmJ}). Using $\xi_{{\rm min}} = 1
- \varepsilon$ with $\varepsilon \ll 1$ and
\begin{equation}
\label{alphaoverK1}%
\alpha/K_1(\alpha) \approx \alpha^2 ,
\end{equation}
valid for $\alpha \ll 1$, one obtains
\begin{equation}
\label{rectlimit}%
n(E,r) = \frac{A}{4 \pi c r^2}\; Q \left[E_g \left(\frac r c \right)
\right] \frac{dE_g}{dE}
\end{equation}
with
\begin{equation}
\label{IntegralI}%
A = \int \limits_{1-\varepsilon}^1 \xi d\xi
\frac{\alpha^2}{(1-\xi^2)^2}\; \exp \left( -
\frac{\alpha}{\sqrt{1-\xi^2}}\right) = \int \limits_{y_{{\rm
min}}}^\infty y e^{-y} dy =1,
\end{equation}
where $y=\alpha/\sqrt{1-\xi^2}$, $y_{{\rm min}} \rightarrow 0$ with
$\alpha \rightarrow 0$. Thus, the regime (\ref{rectlimit}) is indeed
rectilinear.

Now consider the low-energy regime in Eq.~(\ref{PropagatorJuettmJ}).
At small $E$ and $r \gg l_{\rm diff}(E)$ the diffusive regime, free
of the superluminal-signal problem, is valid, and we demonstrate
that the generalized J\"{u}ttner propagator reproduces the
Syrovatsky propagator, given by Eq.~(\ref{SyrovSolution}). Let us
first determine the range of parameters $\xi$ and $\alpha$ which
corresponds to the diffusion, described by the Syrovatsky solution.
The low $E$ corresponds to small $\lambda (E,t)$, and large enough
$t$ provides $r \gg l_{\rm diff}(E)$. Therefore we have $\alpha=
ct/\sqrt{\lambda} \gg 1$. On the other hand $\sqrt{\lambda(E,t)} >r$
provides the considerable contribution to the observed flux from a
source at distance $r$ and thus we have
\begin{equation}
\xi=\frac{r}{ct} < \frac{\sqrt{\lambda}}{ct}=
\frac{1}{\sqrt{\alpha}} \ll 1 .
\label{eq:xi-diff} %
\end{equation}

With these restrictions on $\alpha$ and $\xi$ we compute now
$P_{gJ}(E,t,r)$ using Eq.~(\ref{PropagatorJuettmJ}). At $\alpha \gg
1$ one has
\begin{equation}
\label{alphaK1gg1}%
\alpha/K_1(\alpha) \approx \sqrt{\frac 2 \pi}
\alpha^{3/2}e^{\alpha}.
\end{equation}
Using $\exp (-\alpha/\sqrt{1-\xi^2}) \approx \exp [-\alpha
(1+\xi^2/2)]$ we obtain from (\ref{PropagatorJuettmJ}) after simple
calculations
\begin{equation}
\label{toSyrovatsky}%
P(E,t,r) = \frac{\theta(ct-r)}{\left[ 4 \pi \lambda(E,t)
\right]^{3/2} }\; \exp\left[- \frac{r^2}{4 \lambda(E,t)} \right],
\end{equation}
which coincides with the Syrovatsky propagator
(\ref{PropagatorSyrovatskyDiff}).

However, it should be demonstrated that the space density of
particles is also reproduced correctly. Putting $P(E,t,r)$ from
(\ref{toSyrovatsky}) into (\ref{PropagatorSolution}) we obtain
indeed the Syrovatsky solution (\ref{SyrovSolution}), but with
$E_g^{\rm min}(r)$ as lower limit of integration instead of $E$ in
the Syrovatsky solution, where $E_g^{\rm min}(r)$ corresponds to the
rectilinear propagation $ct=r$. At very low energy $E_g^{\rm min}(r)
\to E$ (see e.g.\ Fig.~\ref{Fig1}) and both solutions coincide.

As a matter of fact J\"{u}ttner found two versions of
relativization of the Maxwell distribution. His second propagator
results in the generalized propagator given by
\begin{equation}
\label{PropagatorJuettmJ2}%
P_{gJ}(E,t,r) = \frac{\theta(1-\xi)}{4 \pi (c t)^3 }\;
\frac{1}{(1-\xi^2)^{5/2}}\; \frac{\alpha(E,\xi)}{K_2[\alpha(E,\xi)]}\;
\exp\left[- \frac{\alpha(E,\xi)}{\sqrt{1-\xi^2}} \right].
\end{equation}
The difference between calculated UHE proton fluxes for both cases
is less than 0.2\%. In the following we discuss the case of
propagator (\ref{PropagatorJuettmJ}).

Therefore, the proposed generalized J\"{u}ttner propagator
(\ref{PropagatorJuettmJ}) has the correct asymptotic behavior
(rectilinear at high energies and diffusive at low energies). It
interpolates between these two regimes without superluminal
velocities. The recent numerical simulations have confirmed the
J\"{u}ttner distribution of relativistic particles
\citep{Dunkel-prl}, which is generalized in our paper for the time
distribution of the charged particles propagating in magnetic
fields.

\section{Propagation of ultra-relativistic particles in magnetic
fields in expanding universe} \label{SectPropagation}

Charged ultra-relativistic particles propagating in the expanding
universe scatter in magnetic fields. Like in the previous section,
where stationary flat space was considered (e.g.\ a galaxy, or
static universe), in the expanding universe there are two extreme
modes for propagation in magnetic field: the (quasi)rectilinear in
the highest energy limit and the diffusive in the low-energy limit.
For these limits we follow the works \citep{BG06,BG07}, where
cosmological effects have been properly included.

For diffusive propagation of particles from a single source located
at the comoving coordinate $\vec{x}_g$ with a generation
function $Q(E,t)$ the diffusion equation has been derived in the
form:
\begin{equation}
\label{DiffuseMF}%
\frac{\partial n}{\partial t} - b(E,t)\; \frac{\partial n}{\partial E}
+ 3 H(t)\; n - \frac{\partial b(E,t)}{\partial E} \; n -
\frac{D(E,t)}{a^2(t)}\; \nabla^2_x n= \frac{Q(E,t)}{a^3(t)}\; \delta^3(\vec{x} -
\vec{x}_g),
\end{equation}
where the cosmological basis $(\vec{x},t)$ is used, $n(E,t,\vec{x})$
is the space density of the particles, $b(E,t) = -dE/dt$ gives the
sum of collisional energy losses, $b_{{\rm int}}(E)$, and adiabatic
energy losses $H(t)E$, $H(t)$ is the Hubble parameter, $D(E,t)$ is a
diffusion coefficient and $a(t)$ is the scaling factor of the
expanding universe.

It was shown that the solution to Eq.~(\ref{DiffuseMF}) reduces to a
quadrature. Introducing the analog of the Syrovatsky variable
\begin{equation}
\label{LambdaExpandUniv}%
\lambda(E,t,t') = \int \limits_{t'}^t dt''\;
\frac{D(\mathcal{E}'',t'')}{a^2(t'')},
\end{equation}
where $\mathcal{E} = E_g(E,t,t')$ is a solution of the
characteristic equation, the solution to Eq.~(\ref{DiffuseMF}) is
given by
\begin{equation}
\label{nExpUniv}%
n(t_0,\vec{x},E) = \int_0^{z_g} dz \left| \frac{dt}{dz} \right|
Q[E_g(E,z),z] \times \frac{\exp \left[- (\vec{x} - \vec{x}_g)^2/4
\lambda(E,z) \right] }{\left[ 4 \pi \lambda(E,z)\right]^{3/2}},
\end{equation}
where $z$ is a redshift, $t_0$ is an age of the universe
corresponding to $z=0$, $z_g$ is the maximum redshift at which a
source is present, $dt/dz = - 1/[(1+z)H(z)]$, and $E_g(E,z)$ is the
generation energy in the source.

The equation describing the rectilinear propagation reads
\citep{BG06}:
\begin{equation}
\label{RectilinearExpandUniv}%
\frac{\partial n}{\partial t} - b(E,t)\frac{\partial n}{\partial E}
+ 3 H(t)\; n - \frac{\partial b(E,t)}{\partial E} \; n + \frac{c
\vec{e}}{a(t)}\; \frac{\partial n}{\partial \vec{x}} =
\frac{Q(E,t)}{a^3(t)}\; \delta^3(\vec{x} - \vec{x}_g),
\end{equation}
where $\vec{e}$ is a unit vector in the direction of propagation.
The solution of Eq.~(\ref{RectilinearExpandUniv}) is given by
\begin{equation}
\label{SolutRectilinearExpandUniv}%
n(t_0,E) = \frac{Q(E_g,t_g)}{4 \pi c\; x_s^2 \;(1+z_g)} \frac{dE_g}{dE},
\end{equation}
where $x_s$ is the comoving distance to a source.

To use the generalized J\"{u}ttner distribution for the expanding
universe we must modify the parameters $\alpha(t)$ and $\xi(t)$
introduced in Section \ref{sec:Approach}.

To provide cosmologically invariant character of the parameters it
is natural to substitute $c t$ by the comoving length of the
particle trajectory $\zeta(t)$:
\begin{equation}
\label{xdistance}%
\zeta(t) = \int_t^{t_0} \frac{c dt}{a(t)} = \frac{c}{H_0}
\int_0^{z_g} \frac{dz}{\sqrt{\Omega_m(1+z)^3 + \Omega_\Lambda}}.
\end{equation}
Now we obtain $\alpha(t)$ from the original quantity $c^2 \tau/2 D$
in the J\"{u}ttner distribution \citep{Dunkelv2}, where $\tau$ is
propagation time, as
\begin{equation}
\label{getalpha}%
\frac{c^2 \tau}{2 D} = \frac{c^2 \tau^2}{2 D \tau} \rightarrow
\frac{\zeta^2(t)}{2 \lambda(E,t)} \equiv \alpha(E,t),
\end{equation}
and use a new variable $\xi$ instead of $t$:
\begin{equation}
\label{xit}%
\xi(t) = x_s/\zeta(t).
\end{equation}
For rectilinear propagation $\xi(t)=1$ holds as before.

In terms of the variables $\alpha$ and $\xi$ defined by
Eqs.~(\ref{getalpha}), (\ref{xit}) the propagator $P_J(E,t,x_s)$ and
particle space density $n(E,x_s)$ have the form similar to
Eqs.~(\ref{PropagatorJuettmJ}) and (\ref{nJuettmJ}):
\begin{equation}
\label{PropJuettExp}%
P_{gJ}(E,t,x_s) = \theta(1-\xi)\; \frac{\xi^3}{x_s^3 (1-\xi^2)^2 }\;
\frac{\alpha}{4 \pi K_1(\alpha)} \exp\left(-
\frac{\alpha}{\sqrt{1-\xi^2}} \right).
\end{equation}

\begin{equation}
\label{nJuettExp}%
n(E,x_s) = \frac 1{4 \pi c x_s^2} \int \limits_{\xi_{{\rm min}}}^1
\frac{\xi d\xi}{1+z(\xi)}\; \frac{Q[E_g(E,\xi)]}{(1-\xi^2)^2}\;
\frac{\alpha}{K_1(\alpha)} \exp\left(- \frac{\alpha}{\sqrt{1-\xi^2}}
\right) \frac{dE_g}{dE}.
\end{equation}

The rectilinear case corresponds to $\xi \approx 1$ and $\alpha \ll
1$. Integral (\ref{nJuettExp}) can be evaluated using
$\alpha/K_1(\alpha) \approx \alpha^2$ at $\alpha \ll 1$ and
integrating over $\xi$ in a narrow range over $\xi \approx 1$. One
obtains Eq.~(\ref{SolutRectilinearExpandUniv}) as it is expected.

The diffusive regime corresponds to $\alpha \gg 1$ and $\xi <1$.
Using $\alpha/K_1(\alpha)$ as given by Eq.~(\ref{alphaK1gg1}) one
obtains for the propagator the correct diffusion expression
\begin{equation}
\label{PropGetApprox}%
P_{gJ}(E,x) = \frac{\exp\left[-x^2/4 \lambda(E,t)\right]}{[4 \pi
\lambda(E,t)]^{3/2}}.
\end{equation}

Thus, the generalized J\"{u}ttner propagator
(\ref{PropagatorJuettmJ}) has the correct asymptotic form in
rectilinear and diffusive regimes again, providing the interpolation
between them without superluminal velocities.

\section{UHECR diffuse spectra in the generalized J\"{u}ttner
approximation}
\label{comparison} %
In this section we calculate the diffuse spectra of UHE protons
propagating in magnetic fields in the expanding universe and compare
them with the calculations of \citep{BG07}, where such spectra have
been calculated for diffusive propagation with transition to
rectilinear propagation at highest energies. The proposed
generalized J\"{u}ttner approximation allows both for diffusive and
rectilinear propagation at low and high energies, respectively, and
provides the interpolation between them based on mathematical
analogy between Maxwellian and diffusion distribution (see
\citep{Dunkelv2} and Eqs.~(\ref{eq:Max}),
(\ref{PropagatorSyrovatsky})).

\begin{figure}
\begin{center}
\includegraphics[width=0.49\textwidth,angle=0]{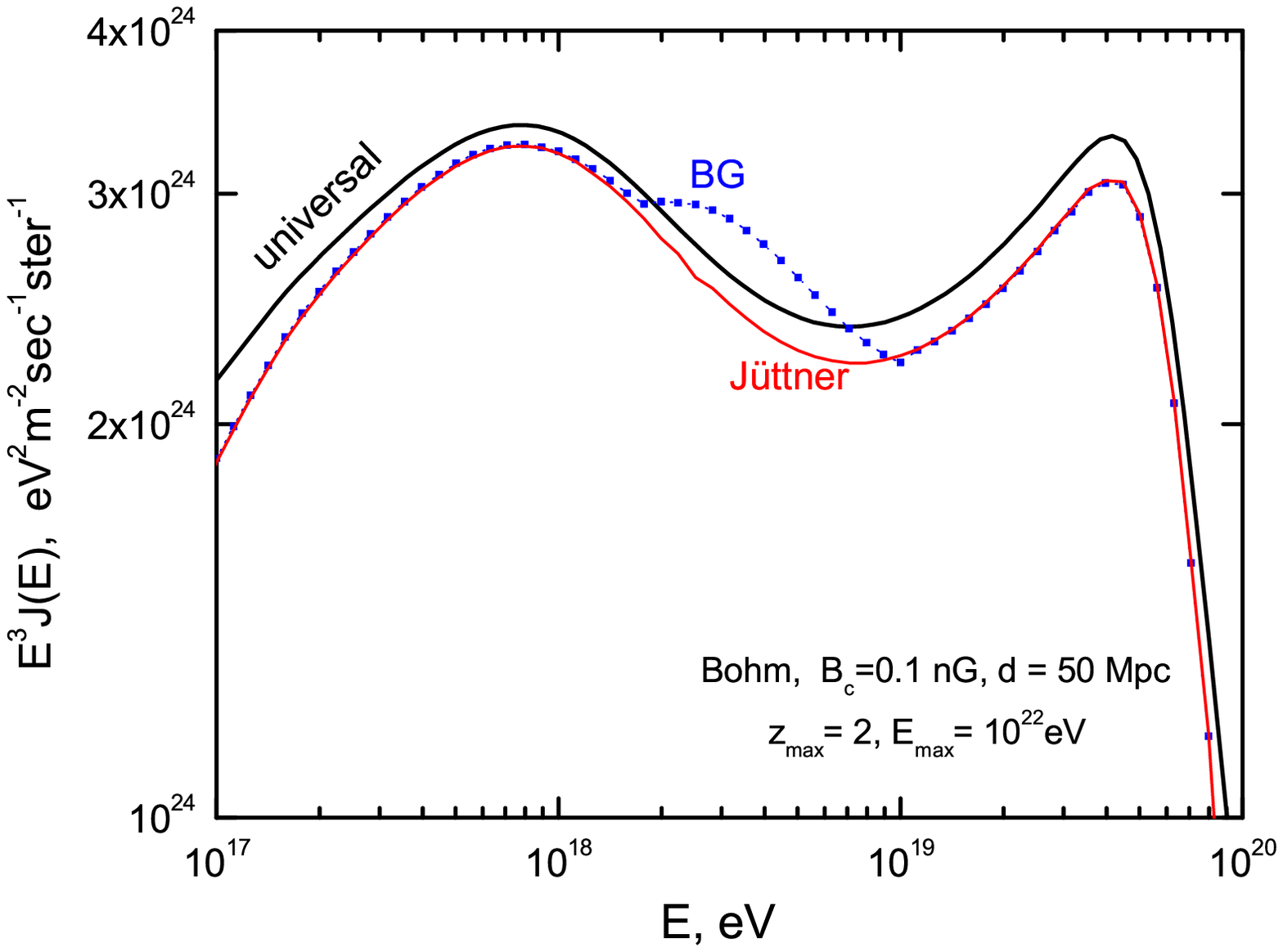}
\includegraphics[width=0.49\textwidth,angle=0]{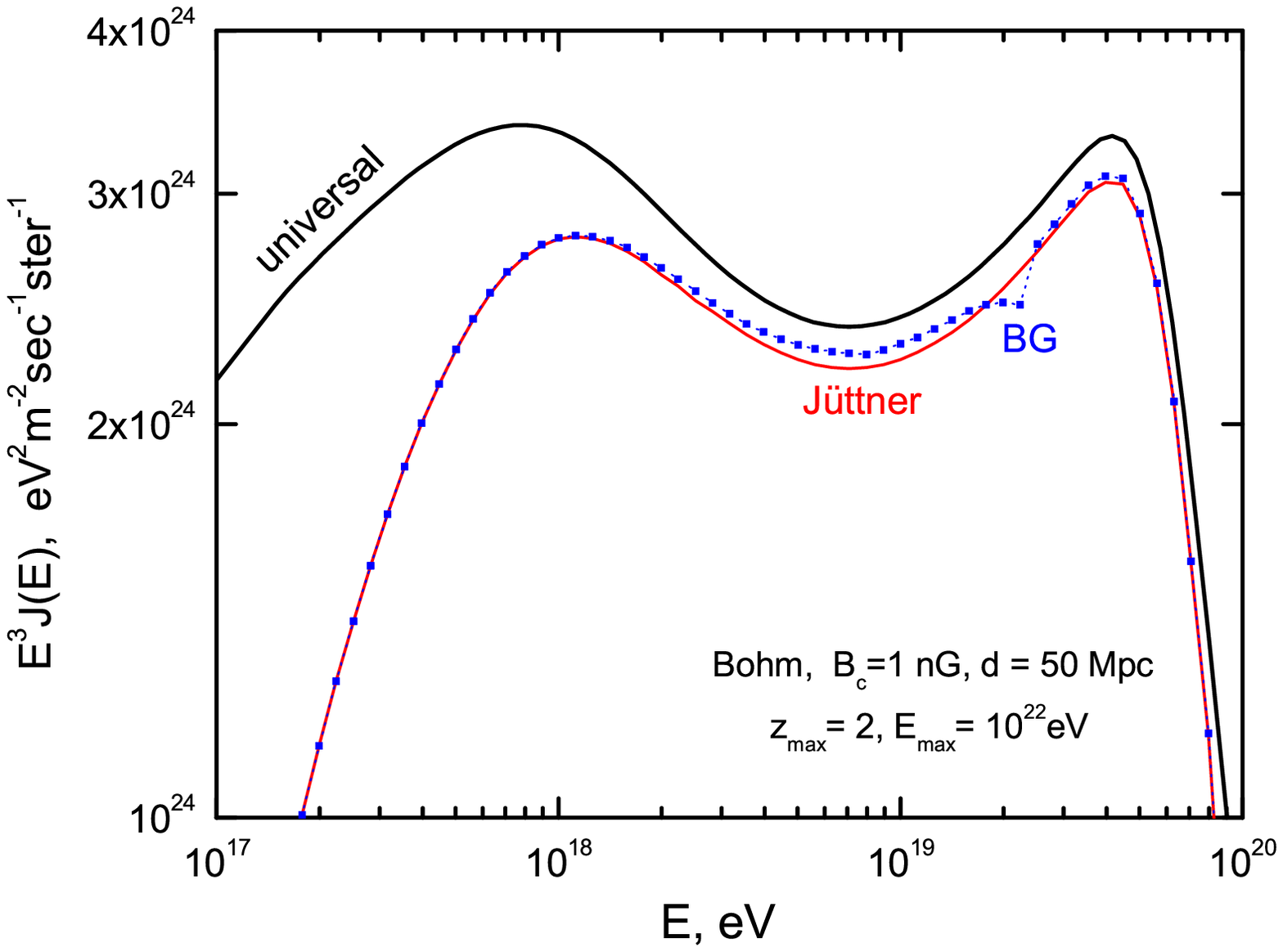}
\end{center}
\caption{
\label{fig2}%
Comparison of the J\"{u}ttner solution with the combined diffusive
and rectilinear solution (BG dotted curves) from work \citep{BG07}.
The left panel shows the case $B_c=0.1$~nG, and the right panel
$B_c=1$~nG, the distance between sources $d=50$~Mpc in both cases
and $\gamma_g=2.7$. The universal spectrum is also presented for
$\gamma_g=2.7$. The features seen in the BG spectra are artifacts
produced by assumption about transition from diffusive to
rectilinear propagation (see text). These features are small: note
the large scale on the ordinate axis. }
\end{figure}

We start our discussion with the universal diffuse spectrum valid
for homogeneous distribution of the sources and for all modes of
propagation. This spectrum can be calculated from the conservation
of number of particles as
\begin{equation}
n(E)dE= \int_{t_{\rm min}}^{t_0} dt\;n_s\;Q[E_g(E,t)]\;dE_g,
\label{eq:universal1}
\end{equation}
where $n_s$ is the source density per unit comoving volume.
According to the propagation theorem \citep{AB04} this spectrum is
independent of mode of particle propagation, and spectrum calculated
for any specific mode of propagation must converge to the universal
one, when distances between sources become less than any
characteristic length involved, e.g.\ the diffusion length.
Therefore, the spectrum (\ref{nJuettExp}) integrated over space
coordinates $x_s$ as
\begin{equation} %
n(E)= \int_0^{\infty} 4\pi x_s^2 dx_s n_s n(E,x_s)
\end{equation} %
must coincide with the universal spectrum (\ref{eq:universal1}).
This fact can be easily demonstrated using
\begin{equation}
\int_0^{\infty} 4\pi x_s^2 \; dx_s \; P_{gJ}(E,t_0,x_s) = 1 \; ,
\label{eq:norm}
\end{equation}
and relations
\begin{equation}
\xi=\frac{x_s}{\zeta} \;\;\; {\rm and}\;\;\;
d\xi=-\frac{x_s}{\zeta^2}\; \frac{c\;dt}{a(t)}. \label{eq:dzeta}
\end{equation}
The calculated universal spectrum is plotted in Fig.~\ref{fig2}.

We calculate now the diffuse flux using the particle density from a
single source given by Eq.~(\ref{nJuettExp}) and assuming the
lattice distribution of the sources in the coordinate space $x$ with
lattice parameter (the source separation) $d$ and a power-law
generation spectrum for a single source,
\begin{equation}
Q_s(E) = \frac{q_0 (\gamma_g - 2)}{E_0^2} \left( \frac{E}{E_0}
\right)^{-\gamma_g},
\end{equation} %
where $E_0$ is the normalizing energy, for which we will use $1
\times 10^{18}$ eV and $q_0$ has a physical meaning of a source
luminosity in protons with energies $E \geq E_0$, $L_p(\geq E_0)$.
The corresponding emissivity $\mathcal{L}_0 = q_0/d^3$, i.e.\ the
energy production rate in particles with $E \geq E_0$ per unit
comoving volume, may be used to fit the observed spectrum by the
calculated one.
\begin{figure}
\begin{center}
\includegraphics[width=0.49\textwidth,angle=0]{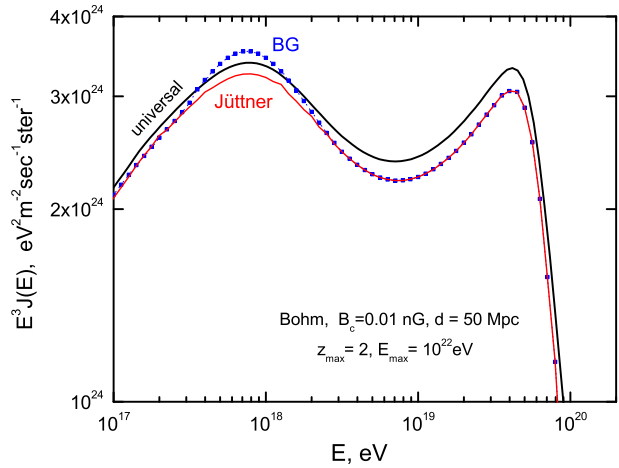}
\end{center}
\caption{
\label{fig3}%
The same as in Fig.~\ref{fig2} for $B_c=0.01$~nG.}
\end{figure}
Using Eq.~(\ref{nJuettExp}) we obtain the diffuse spectrum as
\begin{equation} %
 J_p(E)= \frac{c}{4 \pi H_0} \ \frac{q_0
(\gamma_g-2)}{E_0^2} \sum_s \frac 1{4 \pi c x_s^2} \int
\limits_{\xi_{{\rm min}}}^1 \frac{\xi_s d\xi_s}{1+z(\xi_s)}\;
\frac{[E_g(E,\xi_s)]^{-\gamma_g}}{(1-\xi_s^2)^2}\;
\frac{\alpha}{K_1(\alpha)}\; \exp\left(-
\frac{\alpha}{\sqrt{1-\xi_s^2}} \right) \frac{dE_g}{dE}.
\label{eq:diffuse}
\end{equation} %

For the aim of comparison we use in calculations the same magnetic
field and its evolution with redshift $z$ as in the paper
\citep{BG07}, namely we parametrize the evolution of magnetic
configuration $(l_c,B_c)$ as
\[
l_c(z) = l_c/(1+z), \ \ \   B_c(z)= B_c \ (1+z)^{2-m},
\]
where factor $(1+z)^2$ describes the diminishing of the magnetic
field with time due to magnetic flux conservation and $(1+z)^{-m}$~
due to MHD amplification of the field. The critical energy $E_c(z)$
found from $r_L(E) = l_c(z)$ is given by
\[
E_c(z)=0.93 \times 10^{18}\ (1+z)^{1-m}\ \frac{B_c}{1~\mbox{nG}}
\]
for $l_c = 1$~Mpc. The maximum redshift used in the calculations is
$z_{\rm max}=2$.

In Fig.~\ref{fig2} we compare the generalized J\"{u}ttner solution
with the combined diffusion-rectilinear solution from \citep{BG07}.
Anticipating the present paper, the sewing of the two solutions,
diffusive and rectilinear, has been done in \citep{BG07}
intentionally under over-simplified assumption that for each
distance to the source $x_g$ the transition occurs at the fixed
energy $E_{\rm tr}(x_g)$, where the calculated density of particles
$n(E,x)$ for diffusive and rectilinear propagation are equal. This
was done to mark the energy (or energy region) of transition in the
diffuse spectrum. In our earlier paper \citep{AB05} the transition
was made smoothly. In Fig.~\ref{fig2} one can see that transition
in the generalized J\"{u}ttner solution occurs very smoothly and the
feature in \citep{BG07} solution is useful to indicate the
transition.
\begin{figure}
\begin{center}
\includegraphics[width=0.49\textwidth,angle=0]{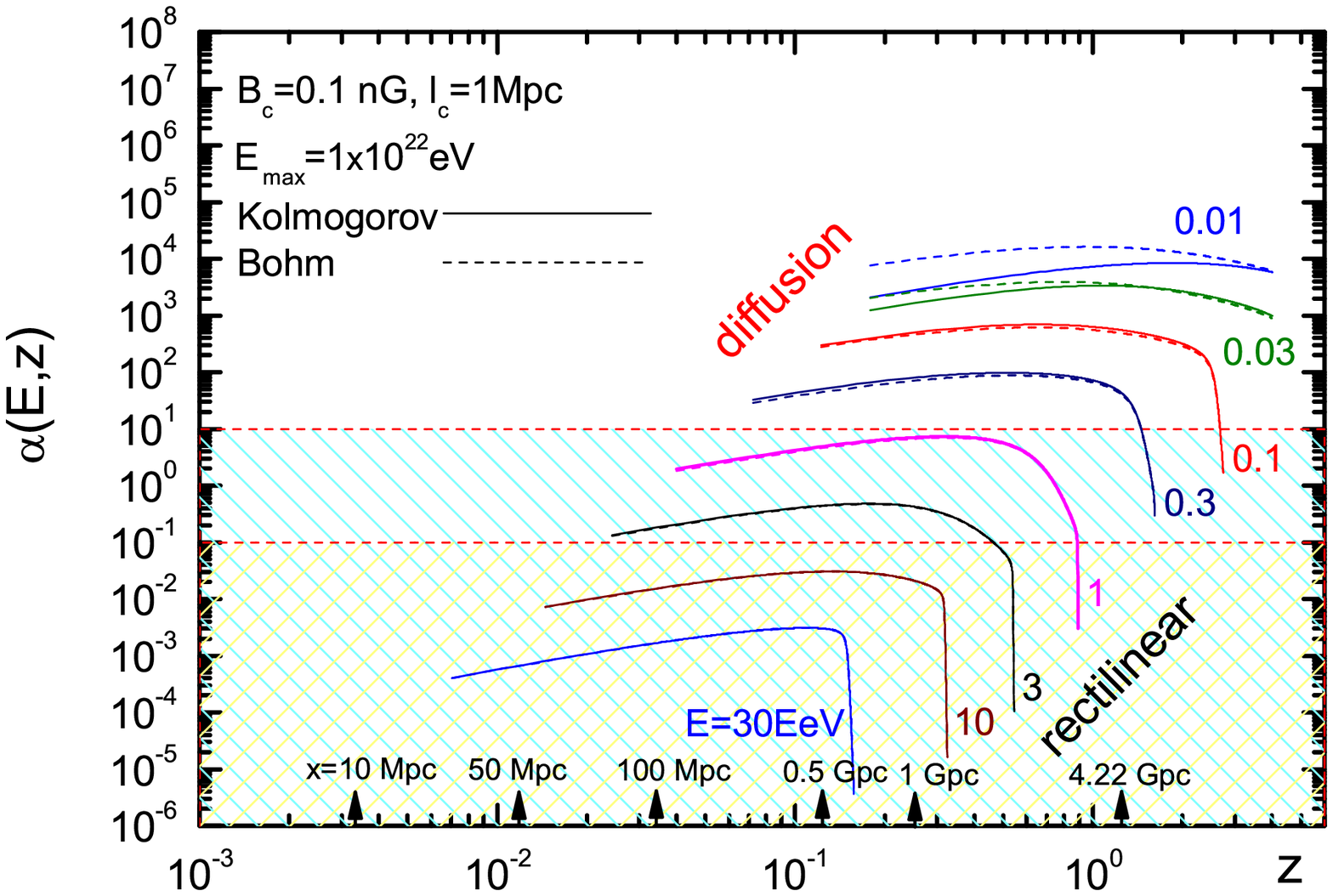}
\end{center}
\caption{
\label{fig4}%
Parameter $\alpha(E,z)$ as a function of
redshift $z$ for the magnetic field configuration $(B_c,l_c)=$
(0.1~nG, 1~Mpc). The observed energies in EeV are shown
at the evolutionary curves. Parameter $\alpha$ changes along the
particle energy trajectory $E_g= {\mathcal E}(E,z)$. The mode of
space propagation (diffusive, intermediate, rectilinear) is changing
accordingly. The regions of propagation correspond to diffusion
($\alpha \geq 10$), rectilinear ($\alpha \leq 0.1$) and intermediate
$(0.1 \leq \alpha \leq 10)$. Along each trajectory the energy $E_g$
increases and $\alpha$ typically decreases due to increasing of
$\lambda (E,E_g)$. }
\end{figure}
As was explained in section \ref{sec:Approach}, the various regimes
of propagation are defined mostly by values of parameters $\alpha$,
in particular, $\alpha \ll 1$ corresponds to rectilinear propagation
and $\alpha \gg 1$ to diffusion. For the expanding universe
$\alpha(E,t)$ is given by Eq.~({\ref{getalpha}).

In Fig.~\ref{fig4} the evolution of $\alpha(E,z)$ is presented as
function of redshift $z$, along the energy trajectories
$E_g={\mathcal E}(E,z)$, where E is the observed energy. The
evolution is shown for different $E$ and magnetic field
configuration $(B_c,l_c)=$(0.1~nG, 1~Mpc).

Fig.~\ref{fig4} illustrates how protons with different observed
energies $E$ propagate in different regimes (diffusive and
rectilinear) and how the regimes are changing. The region with
$\alpha \geq 10$ can be considered as diffusion, $\alpha \leq 0.1$
as rectilinear, and $0.1 \leq \alpha \leq 10$ as intermediate. The
protons with energy $E > 1\times 10^{19}$~eV propagate only
rectilinearly, $E \leq 3\times 10^{16}$~eV only diffusively, and
with other energies -- changing a regime of propagation. Equation
for particle space density (\ref{nJuettExp}) takes into account
automatically this changing of the propagation regimes.
\section{Conclusion}
Diffusion equations are intrinsically non-relativistic and
superluminal velocities appear naturally there. The cardinal
solution of this problem -- the relativistic generalization of the
diffusion equation -- still expects to be found after more than 70
years of unsuccessful attempts.

The {\it phenomenological} approach in its most general form
consists in transition of the diffusive regime to the rectilinear
one in all cases where superluminal propagation appears. We study
here this problem for diffusion of relativistic particles in
magnetic fields. This transition has a problem: there is no extreme
limit in which solution of diffusion equation, e.g.\ in the form of
Eq.~(\ref{SyrovSolution}) or Eq.~(\ref{nExpUniv}), obtains the form
of rectilinear propagation. It requires an intermediate mode of
propagation between two considered extreme regimes. This in
principle follows from the numerical simulations. Diffusive regime
is one where number of particles arriving from the different
directions are equal with great precision and the resulting flux is
given by $\vec{j}=-D \nabla n$. In the rectilinear propagation all
particles arrive from the direction to a source. It is clear that in
intermediate regime the particles are arriving from different
direction with noticeable asymmetry which increases as energy rises,
and analytic solution for such regime does not exists.

Although there is no way to obtain the rectilinear propagation as an
extreme form of solution to diffusion equation, we do know that at
high energy diffusion passes to rectilinear propagation. It follows
from unlimited increase of $D(E)$ with $E$. For example, in the case
of diffusion in turbulent magnetized plasma with maximum scale
$l_c$,~ $D(E) \propto E^2$ at $E > E_c$ (see section
\ref{SuperluminalProblem}). Physically it is clear that very large
diffusion coefficient means rectilinear propagation, but this
argument says nothing about possibility of the formal transition of
the diffusive solution to the rectilinear solution.

Our phenomenological approach for arbitrary mode of propagation is
based on the definition of the propagator in the form of
Eq.~(\ref{PropagatorSolution}):
$$
n(E,r) = \int_0^{\infty} dt\; Q[E_g(E,t),t]\; P(E,t,r)\;
\frac{dE_g}{dE}(E,t),
$$
where $n(E,r)$ is density of particles at distance $r$ from a source
with the rate of particle generation $Q(E_g)$. In principle the
propagator $P(E,t,r)$ must be found as the Green function of the
relativistic diffusion equation, but following the work
\citep{Dunkelv2} we found it from the mathematical analogy with the
Maxwell distribution, given by Eqs.~(\ref{eq:Max}) and
(\ref{PropagatorSyrovatsky}). Then the propagator $P(E,t,r)$ is
found as relativistic J\"{u}ttner propagator, obtained for
relativization of the Maxwell distribution, and expressed in the
terms of diffusion variables \citep{Dunkelv2}. We generalized the
parameters of the J\"{u}ttner distribution
(\ref{PropagatorJuettner}) to meet the more general requirements of
particle propagation: arbitrary dependence of diffusion coefficient
on energy and time and including the energy losses of the particles.
We proved that our generalized J\"{u}ttner propagator
$P_{gJ}(E,t,r)$ given by Eqs.~(\ref{PropagatorJuettmJ}) and
(\ref{PropJuettExp}) satisfies the following conditions:
\begin{itemize}
\item provides in all regimes velocities $v \leq c$,
\item gives in extreme cases (in particular at low and high energies)
       the diffusive and rectilinear propagators,
\item  provides the normalized probability to find a particle
       in a unit volume at distance $r$ from a source at time $t$
       after emission. Equality of this probability integrated over
       space to 1 guarantees the conservation of number of particles,
\item  gives the universal spectrum for homogeneous distribution
       of the sources, as it must be according to the propagation
       theorem \citep{AB04}.
\end{itemize}
In other words, following in principle \citep{Dunkelv2}, we found
the propagator $P_{gJ}(E,t,r)$ from the general requirements listed
above, which include however as the main element the J\"{u}ttner
propagator found from the mathematical analogy between the Maxwell
non-relativistic distribution and diffusion equation
\citep{Dunkel-prd}. This strategy reminds the axiomatic approach to
finding S-matrix in quantum theory, which is built on the basis
of axioms without equations of propagation. Thus, the generalized
J\"{u}ttner propagator gives much more than a simple interpolation
between the diffusive and rectilinear regimes of propagation in
magnetic fields. As one may see from the results, it successfully
describes many features of the unified propagation and may be
thought of as a reasonable approximation to the Green function of
unknown fully relativistic equation for particle propagation, which
includes the diffusion as a low-energy limit.

We have described in this paper the calculation of UHE proton fluxes
in two cases. One is valid for propagation in stationary objects
like e.g.\ galaxies and clusters of galaxies. This solution is
described in section \ref{sec:Approach}. The flux from a single
source at a distance $r$ from an observer is given by
Eq.~(\ref{nJuettmJ}) in terms of the new variable $\xi(t)$ and
parameter $\alpha (E,\xi)$. The diffuse flux can be calculated by
summation over the sources located in vertices of the space
lattice with the source separation $d$.

The second case is given by propagation of UHE protons in the
expanding universe.  The diffuse flux is given by
Eq.~(\ref{eq:diffuse}), where variable $\xi(t)$ and parameter
$\alpha (E,t)$ are defined by the comoving length of a proton
trajectory $\zeta (t)$ (see Eq.~\ref{getalpha}). In these
calculations one does not need to take care of the propagation
regime: it is selected automatically by values of $\xi$ and $\alpha$
in the process of integration.

\section{Acknowledgments} We are grateful to Eugeny Babichev for
discussions. This work is partially funded by the contract ASI-INAF
I/088/06/0 for theoretical studies in High Energy Astrophysics.



\end{document}